\documentclass[12pt]{article}
\usepackage{eurosym}
\usepackage{abstract}
\usepackage{titlesec}
\usepackage{sectsty}
\usepackage{caption}
\usepackage{amssymb}
\usepackage{amsmath}
\usepackage{color}
\usepackage{scalefnt}
\usepackage[super,sort&compress]{natbib}
\usepackage[colorlinks, allcolors=black, urlcolor=blue]{hyperref}
\usepackage{graphicx}
\usepackage[margin=1.35in]{geometry}
\usepackage[section]{placeins}
\usepackage{xr}
\usepackage[normalem]{ulem}

\providecommand{\U}[1]{\protect\rule{.1in}{.1in}}
\titlespacing*{\section}
{0pt}
{0pt}
{8pt}
\titlespacing*{\subsection}
{0pt}
{0pt}
{8pt}

\renewcommand{\arraystretch}{0.6}

\allowdisplaybreaks
\makeatletter
\renewcommand{\maketitle}{
\vspace*{-2em} 
\noindent {\fontsize{16}{18}\selectfont\bfseries \@title \par}
\vspace{0.3em}
\noindent \@author\footnotemark[0] \par
\vspace{0.3em}
\noindent \@date \par
\vspace{1em}
}
\makeatother
\renewcommand{\thefootnote}{\fnsymbol{footnote}}
\begin{document}

\title{Estimating Earth's Temperature Response with Transformed and
Augmented OLS}
\author{Justin Sun \\
{\small Canyon Crest Academy, San Diego, CA 92130, USA }}
\date{}
\maketitle

\begin{abstract}
\ \setlength{\absleftindent}{0pt}

The long-term relationship between radiative forcing and surface temperature
is central to predicting the impacts of climate change. This study employs
multicointegration to characterize this relationship and provides the first
application of the Transformed and Augmented Ordinary Least Squares (TAOLS)
estimator to estimate the multicointegration model. The main objective is
to estimate the Equilibrium Climate Sensitivity (ECS), defined as the global
mean surface temperature increase following a doubling of atmospheric carbon
dioxide. Diagnostic tests reveal that radiative forcing innovations are
strongly non-Gaussian, providing a key motivation for applying
semiparametric TAOLS rather than the parametric Gaussian maximum likelihood
method. TAOLS is also robust to misspecification of the short-run dynamics.
Using the three data pairings of \citet{bruns2020multicointegration}, we
obtain TAOLS estimates of ECS ranging from $1.81^{\circ}$C to
$2.49^{\circ}$C, consistently below their main maximum likelihood estimate
of $2.80^{\circ}$C. Because the model we use is linear and is based on global-mean data, it cannot capture state-dependent feedbacks or regional differences; extending TAOLS along either dimension is a natural next step.
\end{abstract}


\renewcommand{\thefootnote}{\arabic{footnote}}

\section{Introduction}

This paper studies the long-term relationship between radiative forcing and
Earth's surface temperature. Radiative forcing represents the net change in
the energy balance of the Earth system due to external factors and is
measured as the difference between incoming solar radiation and outgoing
thermal energy. This energy imbalance arises primarily from changes in
the concentrations of greenhouse gases such as methane (CH$_{4}$) and
carbon dioxide (CO$_{2}$). Surface temperature refers to the air temperature measured at
approximately two meters above Earth's surface. While surface temperature
responds to changes in radiative forcing, the response is not immediate.
Such a delay is attributed to the oceans' very slow heating rate, as it takes four times as much thermal energy for a unit mass of water to heat up as air. With water covering approximately 71\% of Earth's surface, oceans
absorb between 89\% and 91\% of the excess energy added to the Earth system
due to positive radiative forcing from greenhouse gases, as estimated by von
Schuckmann et al.~\citep{vonschuckmann2023heat} and the Intergovernmental
Panel on Climate Change~\citep[p.~925]{IPCC_2021_WGI}. This leads to a
delayed thermal response that may last from decades to centuries, which
complicates efforts to predict long-term climate outcomes.

This paper focuses on a particular long-term climate prediction: how much the global surface temperature would eventually rise if atmospheric CO$_{2}$ were doubled from pre-industrial levels (circa 1850) and the climate system reached equilibrium. This increase, which accounts for all major climate feedbacks such as water vapor, ice-albedo, and clouds, is defined as the Equilibrium Climate Sensitivity (ECS), a central concept in climate science with practical effects on climate policy. 
Specifically, it refers to the warming from a doubling of pre-industrial CO$%
_{2}$ concentrations (${\sim}$280 parts per million (ppm) to 560 ppm) once
the system reaches equilibrium. For context, the atmospheric CO$_{2}$
concentration was approximately 424 ppm in 2024~\citep{noaa2026co2}.

To estimate the ECS, we need to characterize the long-run relationship
between Earth's surface temperature and radiative forcing. These two time series share the same fundamental statistical characteristics as economic time series. Both exhibit persistent trending behavior, whether deterministic or stochastic, and both are driven by the cumulative effects of structural forces over time. In addition, each represents only a single historical realization over a period that is short relative to the time scales of interest. It is
therefore natural to model both radiative forcing and surface temperature as
integrated processes, as is typically done in econometrics. A process is
called integrated if it can be represented as the cumulative sum of largely
independent shocks; here, cumulative summation is the discrete analog of
integration. Cointegration and multicointegration are the canonical models for this situation: they capture a stable long-run relation between two series although both series may be unpredictable on their own.

The concept of cointegration was introduced and developed by Nobel laureates Robert Engle
and Clive Granger~\citep{EngleGranger1987}. While each time series can drift
or fluctuate independently, cointegration constrains them to move together
in the long run. Cointegration goes hand-in-hand with the error-correction
mechanism (ECM), which connects short-term fluctuations to prior deviations
from the long-term balance. As an example, imagine there are two boats floating in a lake, connected by an elastic rope. Although currents may pull the boats in different directions, the rope’s tension will pull the boats back to equilibrium. Thus, their future movements are shaped by their current separation.

Multicointegration, an extension of cointegration developed by Granger and
Lee~\citep{GrangerLee1989, GrangerLee1990}, builds upon the original concept
by examining how cumulative sums of past imbalances can shape the system's
behavior over longer periods. Going back to the boat example,
multicointegration additionally accounts for how past \textquotedblleft
stretching\textquotedblright\ of the system can affect future movements,
capturing long-term feedbacks.

From the perspective of climate science, cointegration excels at modeling
interactions between radiative forcing and temperature, while
multicointegration is crucial for modeling the delayed effects of
accumulated oceanic heat content on surface warming. While radiative forcing
and temperature may each follow integrated trajectories with local
variations, cointegration can uncover a long-run connection between the two,
indicating they co-move over time. Multicointegration adds to this analysis
by exploring how the cumulative energy stored in the oceans influences
surface temperature. Since the ocean is a slow-responding reservoir of thermal energy, it absorbs heat over decades. The deep ocean then gradually releases this stored energy to the atmosphere over centuries, slowly raising surface temperature. These physical dynamics have a direct formal
counterpart in the econometric framework: \citet{pretis2020equivalence}
shows that a two-component energy balance model (EBM) is mathematically
equivalent to a cointegrated vector autoregression where the climate
feedback parameter is identifiable as a cointegrating coefficient, so that
cointegration is not merely a statistical metaphor imported from economics
but the direct econometric representation of the physical energy balance of
the climate system.

Econometric modeling of climate data offers some advantages over physical
simulation models such as general circulation models (GCMs). As argued by %
\citet{hillebrand2020editorial}, econometric methods ``do not depend on the
accuracy of any complex global climate model'' and instead extract the
long-run signal from the observed statistical properties of the data. In contrast, GCMs calibrate key parameters such as ECS using model simulations. Econometric methods estimate these parameters directly from data, providing
independent, data-driven estimates that complement GCM-based projections
without imposing all of their structural assumptions.

Among existing econometric approaches, the Maximum Likelihood Estimation
(MLE) of Johansen~\citep{johansen1992representation} within an ECM framework
has been widely used to estimate the long-run relationship among climate
time series. For instance, Bruns, Csereklyei, and Stern~%
\citep{bruns2020multicointegration} (hereafter BCS~(2020)) apply MLE within
a multicointegrating model. Although MLE is asymptotically efficient under correct specification, it has two key limitations in our context. First, it imposes a Gaussian distribution on the model innovations. This assumption is likely to be violated when the data contain large non-Gaussian spikes, such as those induced by major volcanic eruptions. Second, it requires correct specification of the short-run VAR dynamics, which is often practically infeasible. This study applies the
Transformed and Augmented Ordinary Least Squares (TAOLS) method of Sun,
Phillips, and Kheifets~\citep{SPK2025} (hereafter SPK~(2025)). TAOLS is more
robust than MLE because it does not impose Gaussianity on the model
innovations and does not require precise specification of the short-run VAR
dynamics. By applying TAOLS to the same dataset from BCS~(2020), this paper
provides a direct comparison of TAOLS and MLE estimates of ECS.

We find that the TAOLS estimates of ECS are consistently lower than the
corresponding MLE estimates reported in BCS~(2020), with each MLE ECS
estimate exceeding the upper bound of the corresponding 95\% TAOLS
confidence interval. Quantitatively, while the MLE yields a point ECS estimate of $%
2.80^{\circ}$C for a focal case, the TAOLS method produces a range of $%
1.81^{\circ}$C to $2.49^{\circ}$C. The precise value of ECS remains a
subject of extensive debate within the climate research community (cf.\ Roe
and Baker~\citep{RoeBaker2007}). Some studies propose an ECS below $2^{\circ}
$C (e.g., Lewis and Curry~\citep{lewis2015implications}), whereas others
indicate values exceeding $5^{\circ}$C (e.g., Bjordal et al.~%
\citep{bjordal2020}). In this context, by applying a more robust estimator
to the same dataset as BCS~(2020), this study contributes to this discourse
by providing empirical evidence for a moderate ECS.

\textit{Related Literature.} Several broad methodological approaches have been developed to estimate ECS.  EBMs, originating with Budyko~%
\citep{budyko1969effect} and Sellers~\citep{sellers1969global}, relate
global-mean temperature change to radiative forcing through a scalar
feedback parameter; \citet{gregory2002new} show how this parameter can be
inferred from GCM regression experiments. \citet{lewis2015implications}
apply the EBM framework to the AR5 observational record and obtain an ECS of
approximately $1.64^{\circ}$C, a low estimate attributable to the ``pattern
effect''~\citep{IPCC_2021_WGI}: the effective feedback parameter changes as
ocean heat uptake patterns evolve, so short-record EBM estimates tend to
understate the true equilibrium response. By contrast, \citet{bjordal2020}
show, using cloud-resolving model simulations, that state-dependent cloud
feedbacks could push ECS above $5^{\circ}$C. The IPCC Sixth Assessment
Report~\citep{IPCC_2021_WGI} synthesizes process-based evidence, the
instrumental record, and paleoclimate reconstructions to argue for a likely
range of $2.5$--$4.0^{\circ}$C. In the econometrics literature, %
\citet{kaufmann1997human} apply cointegration techniques to hemispheric
temperature relations; \citet{kaufmann2013does} provide econometric support
for the $I(1)$ characterization of global temperature. As noted above, %
\citet{pretis2020equivalence} validates the multicointegration framework of %
\citet{bruns2020multicointegration} by establishing the formal equivalence
between EBMs and cointegrated vector autoregressions. %
\citet{phillips2020transient} employ dynamic panel cointegration to estimate
the transient climate sensitivity from station-level data. The
multicointegration approach of BCS~(2020), to which we apply TAOLS, directly
models the accumulation of ocean heat content and estimates the long-run
equilibrium response, thereby avoiding the transient-vs.-equilibrium
conflation inherent in shorter-run EBM methods.

The remainder of this paper is organized as follows. The next section lays
out the theoretical framework, followed by the TAOLS methodology. The
empirical study is then presented, and the final section provides concluding
remarks and discusses future research directions.

\section{Theoretical Framework: Cointegration and Multicointegration\label%
{Sec:TheoreticalFramework}}

\subsection{Cointegration}

We now formalize the statistical framework sketched in the introduction,
beginning with the definition of integrated processes. In the simplest
setting, a time series $\left\{ x_{t}\right\} $ is integrated if it can be
represented as the sum of an independently and identically distributed
(i.i.d.) sequence $\left\{ u_{t}: \mathrm{E}u_{t}=0 \right\} $ 
\begin{equation*}
x_{t}=\sum_{s=1}^{t}u_{s}
\end{equation*}
for all $t=1,\ldots,T$, where $T$ is the length of the time series. In this
formulation, $x_{t}$ represents a \textquotedblleft random
walk,\textquotedblright\ where each new value is the accumulation of all
previous shocks $\left\{ u_{t} \right\}.$ It is clear that the variance of $%
x_{t}$ grows with $t,$ and the process is hence non-stationary. Intuitively,
the process does not revert to a fixed mean; instead, it tends to wander
away from its current position, as the impact of every historical shock is
permanent.

More generally, the zero-mean shocks $\left\{ u_{t}\right\} $ can be
correlated over time, in which case we obtain a general integrated process.
Since there is only one summation involved in the definition of $x_t$, such
a process is called integrated of order one, and we denote this as $%
x_{t}\sim I(1)$. An $I(1)$ process is also referred to as a unit root
process, as its autoregressive representation contains a root equal to one.
On the other hand, we write $u_{t}\sim I(0)$ to signify that it is not
integrated. An $I(0)$ process has a fixed mean and variance and is called
stationary; it tends to revert to its mean over time.

If two time series are individually $I(1)$ processes, but their linear
combination is an $I(0)$ process, then they are cointegrated. As an example,
consider the time series of personal income ($x_{t}$) and consumption
expenditure ($y_{t}$). Both series are affected by stochastic shocks, making them non-stationary. However, they remain tied together by a long-run budget constraint. If
there exists a coefficient $\beta$ such that $y_{t}-\beta x_{t}$ is
stationary, then the two series are cointegrated. This directly implies a
long-run equilibrium relationship despite short-term fluctuations: 
\begin{equation*}
y_{t}=\beta x_{t}+e_{t}
\end{equation*}
where $e_{t}$ is a stationary $I(0)$ error term with a constant mean and
finite variance.

For example, in climate science, radiative forcing ($f_t$) and surface temperature ($s_t$) exhibit similar statistical characteristics. Increases in greenhouse gases and other forcing agents alter the Earth's energy balance, causing radiative forcing to behave as a non-stationary $I(1)$ process. Surface temperature adjusts to these changes over time and likewise exhibits non-stationary $I(1)$ behavior. If a coefficient $\lambda$ exists such that 
\begin{equation*}
q_{t}=f_{t}-\lambda s_{t}
\end{equation*}
is stationary, then $f_{t}$ and $s_{t}$ are cointegrated. Here, $\lambda$
plays the role of the cointegrating coefficient $\beta$ in the general case,
and $q_{t}$ represents the equilibrium error, interpreted physically as the
instantaneous change in Earth's heat content (i.e., the net energy absorbed
or released by the climate system). This cointegrating relationship suggests
that radiative forcing imposes a long-term constraint on surface temperature.

An appealing characteristic of cointegration is its robustness to short-lived noise and temporary deviations from the long-run relationship. For example, aerosol forcing may be unusually high in a particular year, temporarily depressing the surface temperature signal, $s_{t}$. Even so, the underlying relationship with radiative forcing, $f_{t}$, may remain intact. As the temporary disturbance fades, surface temperature converges back toward its long-run equilibrium with radiative forcing. 

\subsection{Multicointegration}

Multicointegration goes beyond cointegration by incorporating cumulative
sums of the equilibrium errors into a second layer of long-term dependence.
Equilibrium errors measure the short-term deviations of the variables from
their long-term cointegration relationships. As an example, consider the
previous analogy with income and expenditure. Let us assume that income and
spending are cointegrated. In a real-world scenario, debt and savings also
exist and they represent the cumulative sum of past deviations between
income and spending. For example, given that a household consistently spends more than its income, it will naturally accumulate debt. Conversely, if it consistently spends less, it will accumulate savings. These accumulated stocks will affect future decisions regarding income and spending. Since the system is governed by both the contemporaneous flow relationship and the cumulative effects of past imbalances, it exhibits multicointegration. 

In the climate domain, multicointegration manifests when the cumulative heat
content ($Q_{t}=\sum_{\tau =1}^{t}q_{\tau }$), derived from the
cointegrating relationship $f_{t}=\lambda s_{t}+q_{t}$, establishes its own
equilibrium with surface temperature ($s_{t}$). Here, $Q_{t}$ represents the
total cumulative energy imbalance of the climate system---encompassing the
oceans (the dominant reservoir), land, ice sheets, and the atmosphere.
Following BCS~(2020), we use ocean heat content as the empirical proxy for
$Q_{t}$, since the oceans absorb the overwhelming majority of the accumulated
imbalance; we acknowledge this is an approximation, as other reservoirs also
contribute to the total energy stored. This stored energy influences surface
temperature through processes such as ocean heat uptake and release,
spanning time periods far longer than regular weather cycles. Therefore,
there may also be a long-run relationship between $Q_{t}$ and $s_{t}$.
Define 
\begin{equation*}
v_{t}=Q_{t}-\phi s_{t}.
\end{equation*}%
Since $Q_{t}$ and $s_{t}$ are both $I(1)$ processes, if there exists a
constant $\phi $ such that $v_{t}$ is stationary, then by definition, $Q_{t}$
and $s_{t}$ are cointegrated. In this case, a multicointegrated system is
formed with two distinct equilibrium relationships: one between $f_{t}$ and $%
s_{t}$, and another between their cumulative equilibrium error ($Q_{t}$) and 
$s_{t}$.

The importance of this second layer of cointegration is particularly evident
in the climate context. For example, a permanent rise in radiative forcing
due to industrial emissions could increase $f_{t}$ over decades and drive $%
s_{t}$ upward while also building up $Q_{t}$ in the deep ocean. Over
centuries, this stored oceanic heat is gradually transferred to the
atmosphere, amplifying $s_{t}$ and creating a long-run feedback that
multicointegration captures.

BCS~(2020) apply the multicointegration framework within a parametric ECM
using the MLE method of Johansen~\citep{johansen1992representation}. Such an
approach is effective at capturing multiple long-run equilibria. However, it
requires explicit parameterization of short-run dynamics. For example, it
has to specify how the change in surface temperature $\Delta s_{t}$ this
year is related to $q_{t-2}$, the heat content flux two years earlier.
Furthermore, it imposes Gaussianity on the VAR innovations, which is not
plausible in our study, as we show in the empirical section.

\section{Methodology: TAOLS Approach \label{Sec:TAOLS}}

The TAOLS method is a semiparametric estimator designed to estimate
multicointegrating relationships with greater flexibility and reduced
computational burden compared to the MLE method of Johansen~%
\citep{johansen1992representation}. We begin with the foundational
cointegrating relationship: 
\begin{equation}
f_{t}=\lambda s_{t}+q_{t} ,  \label{eq:fscoint}
\end{equation}
where $f_{t}$ is radiative forcing (in watts per square meter), $s_{t}$ is
surface temperature (in degrees Celsius), $\lambda$ is the cointegrating
coefficient, and $q_{t}$ is the stationary error representing heat content
flux. Cumulative sums are then defined as 
\begin{equation*}
F_{t}=\sum_{\tau=1}^{t}f_{\tau},\quad S_{t}=\sum_{\tau=1}^{t}s_{\tau},\quad%
\text{and }Q_{t}=\sum_{\tau=1}^{t}q_{\tau}.
\end{equation*}
Taking cumulative sums of equation~\eqref{eq:fscoint} and substituting $%
Q_{t}=\phi s_{t}+v_{t}$, we obtain the multicointegrating regression: 
\begin{equation*}
F_{t}=\lambda S_{t}+Q_{t}=\lambda S_{t}+\phi s_{t}+v_{t}.
\end{equation*}
Here, $\lambda S_{t}$ reflects the cointegration between cumulative forcing
and cumulative temperature, $\phi s_{t}$ represents the multicointegrating
feedback from stored heat to current temperature, and $v_{t}$ represents
unobserved random fluctuations.

Incorporating a constant and a linear time trend into the basic
multicointegration model, we specify the empirical model as%
\begin{equation}
F_{t}=\gamma+\mu t+S_{t}\lambda+s_{t}\phi+v_{t},
\label{Emp_multicointegration_model}
\end{equation}
where the constant $\gamma$ and linear trend $\mu t$ arise naturally from
the multicointegration structure: including an intercept in the level
equation~\eqref{eq:fscoint}, or equivalently in the lower-tier relation $%
Q_{t}=\phi s_{t}+v_{t}$, generates a linear trend in the cumulative equation
upon summation. The linear trend therefore allows for the possibility of a
deterministic trend without imposing one---if no such trend is present in
the data, the estimated coefficient $\hat{\mu}$ will be indistinguishable
from zero. Just as an intercept is routinely included in a regression to
avoid misspecification, including the linear trend here is the appropriate
default.

Following the framework of SPK~(2025), we implement the TAOLS method through
a three-stage process. First, we augment the model with the first difference 
$\Delta s_{t}$ in order to address endogeneity: long-run correlations
between $v_{t}$ and $\Delta s_{t}$ would otherwise bias the estimates of $%
\lambda$ and $\phi$, since short-term fluctuations---such as a $0.3^{\circ}$%
C El~Ni\~{n}o spike---would contaminate the long-run estimates. The
augmented model is 
\begin{equation}
F_{t}=\gamma +\mu t+S_{t}\lambda +s_{t}\phi +\Delta s_{t}\delta +e_{t}.
\end{equation}

Second, we project all variables onto a set of low-frequency basis
functions. Specifically, we use the following sine functions: 
\begin{equation*}
\left\{ \varphi _{i}\left( r\right) =\sqrt{2}\sin \left((i-\frac{1}{2})\pi
r\right):i=1,\ldots ,K\right\}.
\end{equation*}%
Here $r = t/T \in [0,1]$ is a normalized time index. In principle, any
complete orthogonal basis can be used; the key requirement is that the first 
$K$ basis functions have their energy concentrated at low frequencies, so
that the transformation captures the long-run variation of the underlying
integrated series. Sine functions, cosine functions, and Fourier
(sine--cosine) pairs all satisfy this requirement. Following \citet{PK2024},
we adopt the above orthonormal sine basis.

The corresponding transformed variables are:%
\begin{align}
V_{F,i}& =\frac{1}{\sqrt{T}}\sum_{t=1}^{T}F_{t}\varphi _{i}\left( \frac{t}{T}%
\right) ,\text{ }V_{S,i}=\frac{1}{\sqrt{T}}\sum_{t=1}^{T}S_{t}\varphi
_{i}\left( \frac{t}{T}\right) ,  \notag \\
V_{s,i}& =\frac{1}{\sqrt{T}}\sum_{t=1}^{T}s_{t}\varphi _{i}\left( \frac{t}{T}%
\right) ,\text{ }V_{\Delta s,i}=\frac{1}{\sqrt{T}}\sum_{t=1}^{T}\Delta
s_{t}\varphi _{i}\left( \frac{t}{T}\right) ,\text{ }  \notag \\
V_{\ell ,i}& =\frac{1}{\sqrt{T}}\sum_{t=1}^{T}\left( 1,t\right) \varphi
_{i}\left( \frac{t}{T}\right) ,\text{ }V_{e,i}=\frac{1}{\sqrt{T}}%
\sum_{t=1}^{T}e_{t}\varphi _{i}\left( \frac{t}{T}\right) .
\end{align}

These transformed variables retain the low-frequency components of the time series while filtering out high-frequency variation. The
parameter $K$ denotes the number of basis functions  used in the
transformation. Based  on the transformed variables, we now have 
\begin{equation}
V_{F,i}=V_{\ell ,i}\alpha +V_{S,i}\lambda +V_{s,i}\phi +V_{\Delta s,i}\delta
+V_{e,i}  \label{eq:TA_reg}
\end{equation}%
for $\alpha =\left( \gamma ,\mu \right) ^{\prime }$ and $i=1,2,\ldots ,K.$
This is our transformed and augmented regression model.

Third, we apply OLS to the transformed regression model~\eqref{eq:TA_reg} to
obtain estimates of $\gamma$, $\mu$, $\lambda$, $\phi$, and $\delta$. Unlike
Johansen's MLE, which requires a fully specified ECM and assumes normality,
TAOLS relies on the asymptotic properties of the transformed series,
delivering estimators that are consistent and asymptotically mixed-normal
under general conditions (e.g., weak dependence in $e_{t}$). A further advantage of TAOLS is that standard inference methods, such as the $t$-test, can be readily applied, making statistical inference straightforward.
 For details on the asymptotic theory, see \citet{SPK2025}.

\section{Results \label{Sec:Empirics}}

\subsection{Data Sources}

The empirical study uses data from \citet{bruns2020multicointegration},
which were obtained from multiple sources and span the period 1850--2014,
yielding $T=165$ annual observations, as detailed in Appendix~A of %
\citet{bruns2020multicointegration}. Radiative forcing is computed using an
established formula for seven components: well-mixed greenhouse gases, solar
irradiance, tropospheric sulfate aerosol, black carbon, organic carbon,
ozone, and stratospheric aerosol. For example, the radiative forcing of
carbon dioxide (CO$_{2}$), a critical greenhouse gas in climate research, is
calculated using the logarithmic relationship: 
\begin{equation}
RF_{CO_{2}}=5.35\times \ln \left( \frac{C}{C_{0}}\right) ,  \label{eq:RFCO2}
\end{equation}%
where $C$ is the CO$_{2}$ concentration in ppm, and $C_{0}$ is the
pre-industrial level (1850), approximately 280 ppm. In the above, $5.35$ is
an empirically derived coefficient from \citet{myhre1998new} that converts a
change in CO$_{2}$ concentration into a change in Earth's radiative energy
balance. We consider two versions of aggregate radiative forcing:
full-efficacy radiative forcing (hereafter TotalRF) and partial-efficacy
radiative forcing (hereafter MarvelRF), the latter adjusting the forcing
from ozone, volcanic aerosols, and solar irradiance by a factor of 0.5.

Surface temperature data are sourced from Berkeley Earth~%
\citep{berkeleyearth2025data} and the Hadley Centre/Climatic Research Unit
(HadCRUT)~\citep{hadcrut4data}. Temperatures, reported in degrees Celsius,
are expressed as anomalies relative to the January 1951--December 1980
average.

\begin{figure}[ptb]
\centering
\includegraphics[width=\textwidth]{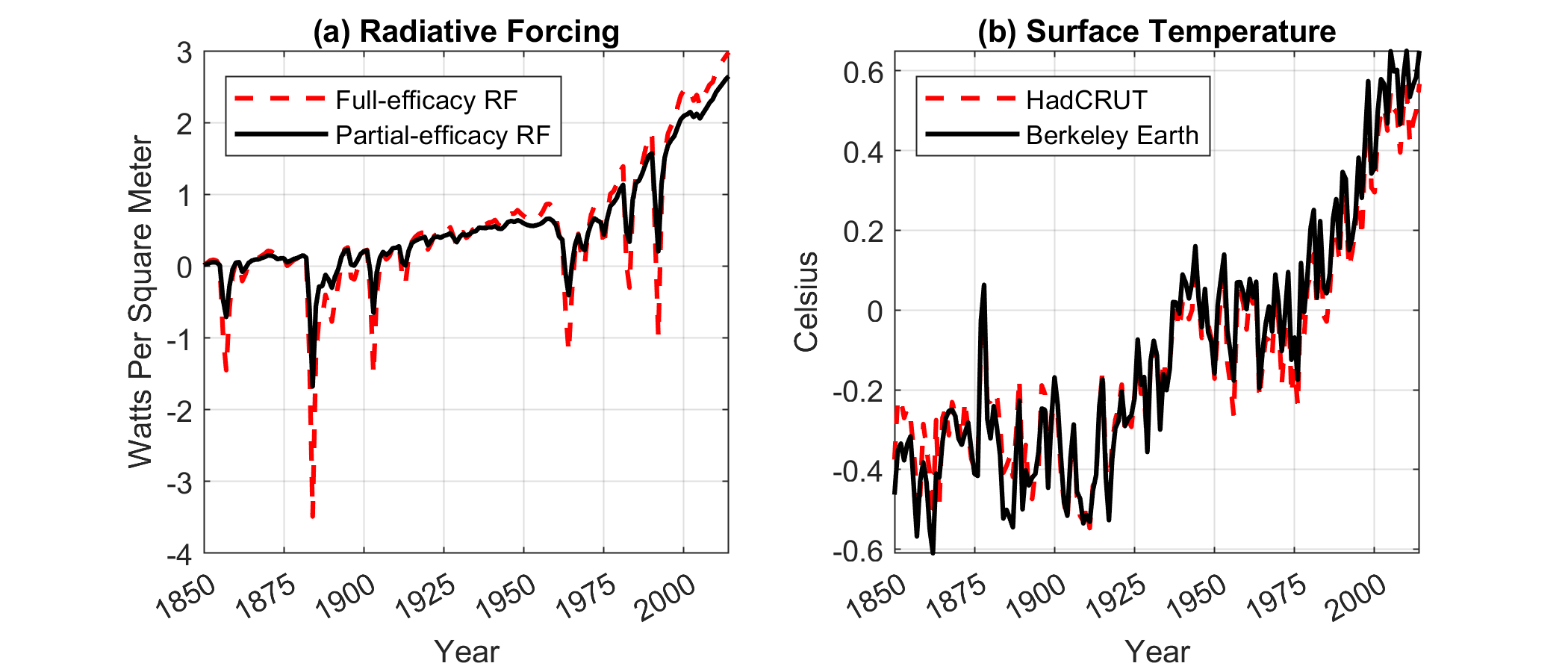}
\caption{Radiative forcing (panel~a) and surface temperature anomaly
(panel~b), 1850--2014. Panel~(a): full-efficacy and partial-efficacy
radiative forcing (W\,m$^{-2}$). Panel~(b): Berkeley Earth and HadCRUT
surface temperature anomalies ($^{\circ}$C, relative to 1951--1980). }
\label{figure:RF_temperature}
\end{figure}

Figure~\ref{figure:RF_temperature} reproduces Figures 1 and 2 of %
\citet{bruns2020multicointegration}, to provide a visual overview of the data.
Panel~(a) plots both radiative forcing series and panel~(b) plots both
surface temperature series. Radiative forcing and surface temperature
display evident co-movement. However, the presence of large spikes induced
by volcanic eruptions suggests that radiative forcing is unlikely to conform
to a normal distribution. Consequently, applying a VAR system under the
normality assumption, as implemented in \citet{bruns2020multicointegration},
may yield unreliable estimates. The differences between the two versions of
each series indicate some uncertainty in their measurement. If these
differences are of a high-frequency nature, then TAOLS is particularly
well-suited, as it does not rely on the high-frequency components of the
underlying time series.

\subsection{Stochastic Trend and Multicointegration Assumptions\label%
{Sec:Diagnostics}}

The TAOLS framework requires that radiative forcing $f_{t}$ and surface
temperature $s_{t}$ each be integrated of order one ($I(1)$) and that the
system be multicointegrated. These properties are supported by formal unit-root tests, physical reasoning, and the system-level rank tests reported by \citet{bruns2020multicointegration}.

\textit{Unit-root $I(1)$ tests.} We apply two complementary unit-root tests
to all four series---TotalRF, MarvelRF, Berkeley Earth temperature, and
HadCRUT temperature---over the full sample 1850--2014 ($T=165$). The
Augmented Dickey--Fuller (ADF) test~\citep{dickey1979distribution} has the
null hypothesis of a unit root; failure to reject is evidence of $I(1)$
behavior. The KPSS test~\citep{kwiatkowski1992testing} has the null
hypothesis of trend stationarity; rejection is evidence of a unit root. A
pattern of ADF non-rejection combined with KPSS rejection thus constitutes
strong evidence in favor of $I(1)$. For the ADF test, we include a constant
(no deterministic trend) and select the lag length by BIC, up to a maximum
of three lags. For the KPSS test, we use an automatic bandwidth selection
procedure and test the null of trend stationarity. Table~\ref{tab:unitroot}
reports the results.

\begin{table}[htbp]
\caption{Unit-Root Test Results (1850--2014, $T=165$)}
\label{tab:unitroot}\renewcommand{\arraystretch}{1.2} \centering
\begin{tabular}{l|cc|cc}
\hline\hline
& \multicolumn{2}{c|}{ADF: Constant} & \multicolumn{2}{c}{KPSS: Trend} \\ 
Series & Stat (lag) & $p$-val & Stat & $p$-val \\ \hline
TotalRF & $-1.653$ (2) & 0.449 & 0.921 & ${\leq}0.010$ \\ 
MarvelRF & $-0.179$ (2) & 0.938 & 1.667 & ${\leq}0.010$ \\ 
Berkeley & $-0.073$ (3) & 0.950 & 1.481 & ${\leq}0.010$ \\ 
HadCRUT & $-0.098$ (3) & 0.947 & 1.635 & ${\leq}0.010$ \\ \hline\hline
\end{tabular}%
\par
\begin{flushleft}
{\footnotesize Notes: ADF $H_0$: unit root; lag by BIC (max 3); 5\% CV $=
-2.87$. KPSS $H_0$: trend stationary; 5\% CV $= 0.146$, 1\% CV $= 0.216$. }
\end{flushleft}
\end{table}

For all four series, the ADF statistic is well above the 5\% critical value
of $-2.87$, so the null of a unit root is not rejected at any conventional
significance level. Symmetrically, all four KPSS statistics far exceed the
1\% critical value of $0.216$, so the null of trend stationarity is strongly
rejected. The two tests agree: every series is $I(1)$.

The $I(1)$ characterization can also be based on physical reasoning and
empirical evidence from the broader econometric literature. Radiative
forcing is driven by the slow accumulation of greenhouse gas concentrations
from persistent anthropogenic sources; surface temperature adjusts to
forcing changes with a multi-decadal lag governed by ocean thermal inertia.
As argued by \citet{bruns2020multicointegration} following %
\citet{kaufmann2010does}, absent radiative forcing the climate system would
be mean-reverting; it is because forcing itself follows a stochastic trend
that temperature does likewise. \citet{kaufmann2013does} directly evaluate
conflicting statistical claims about the integration order of surface
temperature, including arguments in favor of trend-stationary
representations, and conclude that the data cannot reject the unit-root
characterization once the analysis is conducted carefully.

Moreover, \citet{bruns2020multicointegration} note, following %
\citet{kejriwal2013unit}, that a unit-root process is the limiting case of a
trend-stationary process with a structural break every period: with enough
breakpoints any random walk can be approximated by a piecewise linear
trend-stationary process, making the two representations nearly
observationally equivalent in samples of this length. %
\citet{bruns2020multicointegration} conduct Johansen $I(2)$ rank tests on
the identical dataset and sample period, concluding that the two series are
cointegrated and the system is multicointegrated. Since our analysis uses
the same data and sample, these results carry over directly.

\textit{Normality of innovations.} The Jarque--Bera (JB) test~%
\citep{jarque1980efficient} examines whether the skewness and excess
kurtosis of a series are consistent with the Gaussian distribution. Because
the level series are $I(1)$, we fit an AR(3) model, $\Delta y_t = c +
\rho_1\Delta y_{t-1} + \rho_2\Delta y_{t-2} + \rho_3\Delta y_{t-3} +
\varepsilon_t$, to each first-difference series by OLS and apply JB to the
residuals. Removing autocorrelation structure before testing the
distributional assumption provides a more stringent check of Gaussianity
than applying JB directly to the first differences. The JB statistic $\frac{T%
}{6}\bigl[S^{2}+\frac{(\kappa-3)^{2}}{4}\bigr]$ is asymptotically $%
\chi^{2}(2)$ under the null, where $S$ and $\kappa$ denote sample skewness
and kurtosis of the residuals ($T=161$ after three-lag alignment). Table~\ref%
{tab:normality_ar3} reports the results.

\begin{table}[htbp]
\caption{Jarque--Bera Normality Test for AR(3) Residuals of First Differences
}
\label{tab:normality_ar3}\renewcommand{\arraystretch}{1.2} \centering
\begin{tabular}{lcccc}
\hline\hline
Series & Skewness & Excess Kurtosis & JB Stat & $p$-value \\ \hline
$\Delta$TotalRF & $-1.019$ & $\phantom{-}8.131$ & $471.3$ & ${\leq}0.001$ \\ 
$\Delta$MarvelRF & $-0.893$ & $\phantom{-}7.870$ & $436.9$ & ${\leq}0.001$
\\ 
$\Delta$Berkeley & $\phantom{-}0.104$ & $-0.279$ & $\phantom{0}0.8$ & ${\geq}%
0.500$ \\ 
$\Delta$HadCRUT & $\phantom{-}0.129$ & $-0.452$ & $\phantom{0}1.8$ & $0.341$
\\ \hline\hline
\end{tabular}%
\par
\begin{flushleft}
{\footnotesize Notes: AR(3) model $\Delta y_t = c + \rho_1\Delta y_{t-1} +
\rho_2\Delta y_{t-2} + \rho_3\Delta y_{t-3} + \varepsilon_t$, fitted by OLS. 
$T=161$ residuals (1854--2014). $H_{0}$: normality ($\chi^{2}(2)$). $p$%
-values bounded at 0.001/0.500 by MATLAB's \texttt{jbtest}. }
\end{flushleft}
\end{table}

Both radiative forcing series reject normality decisively (JB $>430$, $%
p\leq0.001$), driven by heavy-tailed outliers from major volcanic eruptions
(e.g.\ Krakatoa in 1883, Pinatubo in 1991) that generate large year-on-year
swings in forcing not captured by the linear AR(3) dynamics. By contrast,
both temperature series are consistent with normality ($p>0.34$), reflecting
the smoothing effect of ocean thermal inertia. This non-normality of
radiative forcing innovations invalidates the Gaussian MLE used by %
\citet{bruns2020multicointegration} and provides the key motivation for
TAOLS, which does not impose Gaussianity. Histograms and normal Q-Q plots
are available upon request.

\subsection{Multicointegration and ECS}

Following \citet{bruns2020multicointegration}, we compute ECS from the
estimated cointegrating coefficient $\lambda$ via 
\begin{equation}
\mathrm{ECS}=\frac{5.35\times\ln(2)}{\lambda}=\frac{3.71}{\lambda},
\end{equation}
where $3.71\approx 5.35\times\ln 2$ is the radiative forcing from a doubling
of CO$_{2}$ concentrations (equation~\eqref{eq:RFCO2} with $C/C_{0}=2$).
Although our empirical model uses aggregate radiative forcing, anthropogenic
forcing over the instrumental record is dominated by long-lived greenhouse
gases whose aggregate forcing is approximately proportional to CO$_{2}$
forcing, so the CO$_{2}$-doubling benchmark remains the standard convention.

Our empirical analysis considers three data pairings from %
\citet{bruns2020multicointegration}: Model~I uses full-efficacy radiative
forcing with Berkeley Earth surface temperature; Model~II uses
partial-efficacy radiative forcing with Berkeley Earth surface temperature;
and Model~III uses partial-efficacy radiative forcing with HadCRUT surface
temperature. Strictly speaking, these are three data pairings rather than
three separate models, but we use the same terminology for easy comparison.

\begin{figure}[ptb]
\centering
\includegraphics[width=6.0in]{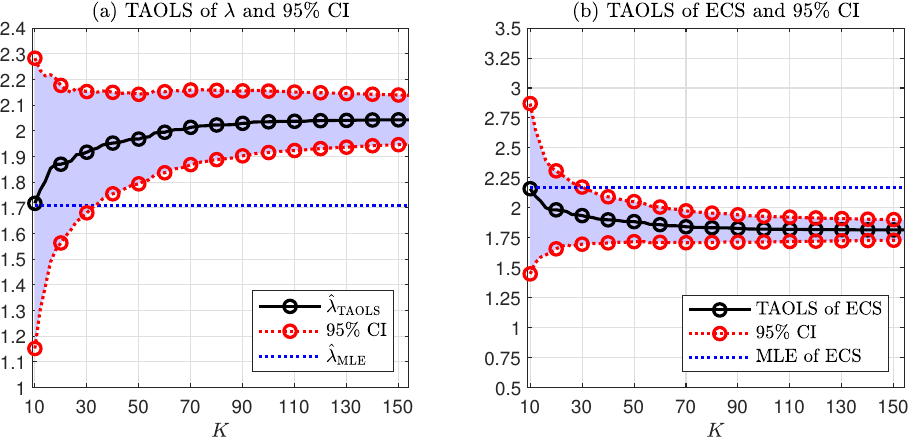}
\caption{TAOLS estimate of $\protect\lambda$ and the implied ECS in the
presence of multicointegration for different values of $K$ using
full-efficacy radiative forcing and surface temperature from Berkeley Earth
(Model~I). The shaded region represents the 95\% confidence interval.}
\label{figure:model_I_lambda_ECS}
\end{figure}

\begin{figure}[ptb]
\centering
\includegraphics[width=6.0in]{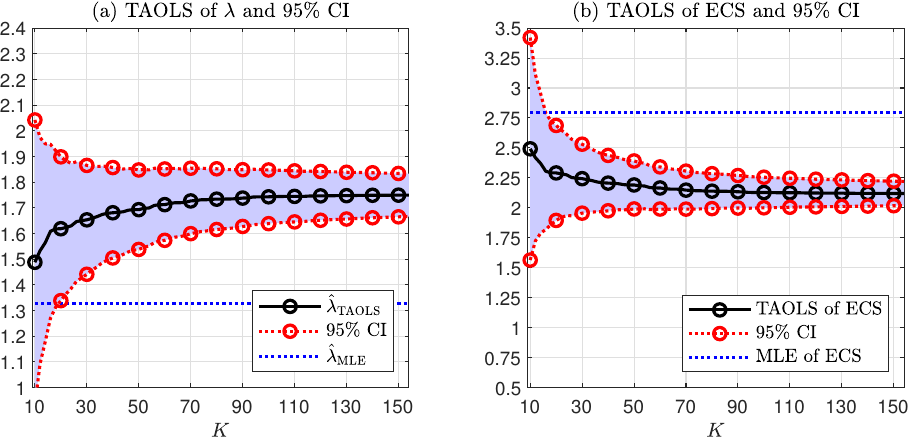}
\caption{TAOLS estimate of $\protect\lambda$ and the implied ECS in the
presence of multicointegration for different values of $K$ using
partial-efficacy radiative forcing and surface temperature from Berkeley
Earth (Model~II). The shaded region represents the 95\% confidence interval.}
\label{figure:model_II_lambda_ECS}
\end{figure}

\begin{figure}[ptb]
\centering
\includegraphics[width=6.0in]{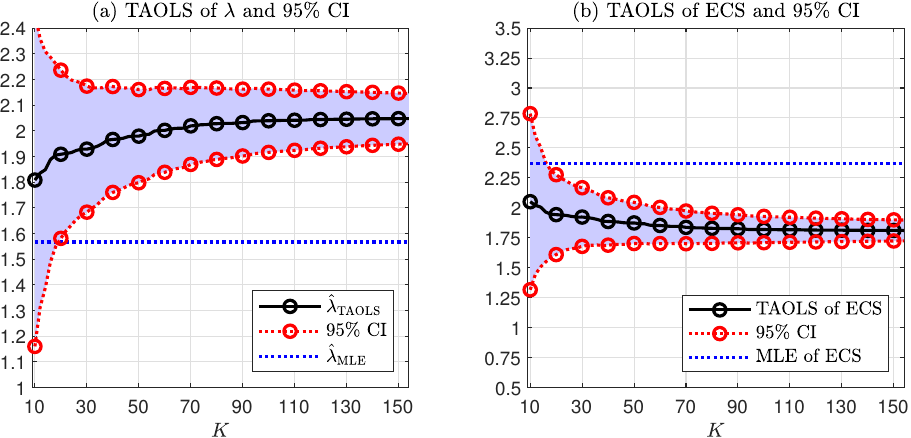}
\caption{TAOLS estimate of $\protect\lambda$ and the implied ECS in the
presence of multicointegration for different values of $K$ using
partial-efficacy radiative forcing and surface temperature from HadCRUT
(Model~III). The shaded region represents the 95\% confidence interval.}
\label{figure:model_III_lambda_ECS}
\end{figure}

Figures~\ref{figure:model_I_lambda_ECS}--\ref{figure:model_III_lambda_ECS}
display the TAOLS estimates of $\lambda$ and the implied ECS for each data
pairing. In Model~I, $\lambda$ ranges from $1.717$ to $2.043$ with a mean of 
$1.991$, yielding ECS estimates in $\left[1.82,2.16\right]^{\circ}$C with a
mean of $1.87^{\circ}$C. For Model~II, $\lambda$ lies in $\left[1.488,1.750%
\right]$ with a mean of $1.709$, corresponding to ECS in $\left[2.12,2.49%
\right]^{\circ}$C with a mean of $2.17^{\circ}$C. In Model~III, $\lambda$
ranges from $1.808$ to $2.048$ with a mean of $2.002$, implying ECS in $%
\left[1.81,2.05\right]^{\circ}$C with a mean of $1.85^{\circ}$C.

These results diverge from those of \citet{bruns2020multicointegration}, who
employed MLE within an ECM framework. Their MLE estimates of $\lambda$ are $%
1.709$, $1.326$, and $1.567$ for Models~I, II, and~III respectively,
implying ECS of $2.17^{\circ}$C, $2.80^{\circ}$C, and $2.37^{\circ}$C.
Across all three specifications, the TAOLS estimates of $\lambda$ exceed the
corresponding MLE values, yielding ECS estimates that range from $%
1.81^{\circ}$C to $2.49^{\circ}$C, all below the corresponding MLE
benchmarks. Compared to \citet{bruns2020multicointegration}, the standard
errors of $\hat{\lambda}_{\mathrm{TAOLS}}$ are generally smaller, producing
narrower confidence intervals. In each specification, $\lambda$ increases
with $K$ but stabilizes around $K=80$, beyond which additional basis
functions have negligible effect on the estimates, a finding consistent with
asymptotic efficiency arguments in \citet{SPK2025}.

A few comments on the choice of $K$ are in order. At the lower end, $K$ has
to exceed the number of regressors in the TAOLS multicointegration
regression (i.e., 5). Setting $K=10$ leaves only $10-5=5$ residual degrees
of freedom, the minimum we regard as adequate for reliable variance
estimation; reducing to $K=5$ would exhaust all degrees of freedom ($5-5=0$%
), making the residual variance estimator undefined. The asymptotic standard
error of each TAOLS coefficient is proportional to $\sqrt{T/K}$; halving $K$
inflates the standard error by a factor of $\sqrt{2}$, which explains the wide
confidence bands visible at $K<30$. At the upper end, $K$ is bounded by the
effective sample size after first-differencing (i.e., $164$). Moreover, as $%
K\to T$, the TAOLS estimator converges to OLS, which suffers from an
asymptotic bias that the sine transformation is designed to remove, so we
stop short of $T$ at $K=154$. While a data-driven choice of $K$ based on an
asymptotic mean-squared-error criterion exists in principle, it cannot be
estimated reliably at sample sizes of this magnitude. More importantly,
quantitatively consistent estimates across the entire feasible range of $K$%
---as we document here---provide stronger evidence of robustness than any
single-$K$ result, which could reflect the particular frequency composition
at that specific value of $K$.

\subsection{Multicointegration and Heat Feedback}

We now examine the estimate of $\phi$. Note that $\phi $ measures the
long-run relationship between the heat content of the system and surface
temperature. At equilibrium, for a given value of $\phi ,$ a rise in surface
temperature by $1^{\circ}$C corresponds to an increase in the system's heat
content of approximately $\phi $ watt-years per square meter (W-yr/m$^{2}$).
For reference, increasing the atmospheric temperature by $1^{\circ}$C
requires an increase in heat content of approximately $5\times10^{21}$
joules---derived from the mass of the atmosphere (${\approx}5.15\times10^{18}
$\,kg, inferred from surface pressure measurements) and the specific heat of
air at constant pressure ($c_p\approx1{,}005$\,J\,kg$^{-1}$\,K$^{-1}$),
giving $5.15\times10^{18}\times1{,}005\approx5.17\times10^{21}$\,J\,K$^{-1}$%
---which translates to 0.31 W-yr/m$^{2}$. Thus, every 0.31 W-yr/m$^{2}$ of
additional heat absorbed by the atmosphere corresponds to $\phi $ W-yr/m$^{2}
$ of the total heat content of the entire system, including the atmosphere,
oceans, and other components. This suggests that approximately $31/\phi $
percent of the total heat content contributes to warming the atmosphere.

For Model~II, Figure~\ref%
{figure:partial_rf_berkeleyt_trendtype_1_mc_phi_percent} shows that the
multicointegration parameter $\phi$ ranges from $10.41$ to $24.51$ across
the $K$ values. To interpret this physically, consider $\phi =20$: a 1$%
^{\circ }$C increase in surface temperature corresponds to an additional 20
W-yr/m$^{2}$ of stored heat, predominantly in the oceans. This implies that $%
31/20\approx 1.5$ percent of the total heat content contributes to
atmospheric warming, with the remainder residing in oceanic or terrestrial
reservoirs. By comparison, \citet{bruns2020multicointegration} report $\phi $ values
of 31 to 41, suggesting an air warming fraction of 1\% or
less. Across all models, the estimated atmospheric share ranges from approximately 0.95\% to 2.98\%. This range closely matches the empirical estimate of roughly 1\% over  1971–2018 \citep[Table~7.1, p.~938]{IPCC_2021_WGI}. The agreement supports the conclusion that most excess heat is stored in the oceans rather than the atmosphere. Note
that $31/\phi$ percent provides an underestimate of the fraction of ocean
heat content directed toward surface warming, since the denominator reflects
total system heat content rather than the ocean component alone; the
difference is small given that the oceans dominate total system heat content.

To illustrate, suppose a sustained radiative forcing increase of 1 W\,m$^{-2}
$ over 10 years accumulates 10 W-yr/m$^{2}$ in $Q_{t}$. With $\phi=20$, this
heat storage corresponds to a surface temperature increase of approximately $%
10/20 = 0.5^{\circ}$C, illustrating the slow feedback from oceanic
reservoirs to atmospheric conditions. This feedback mechanism underscores
the multicointegrating structure's relevance, capturing dynamics beyond the
first layer of cointegration. 
\begin{figure}[ptb]
\centering
\includegraphics[width=6.1056in]{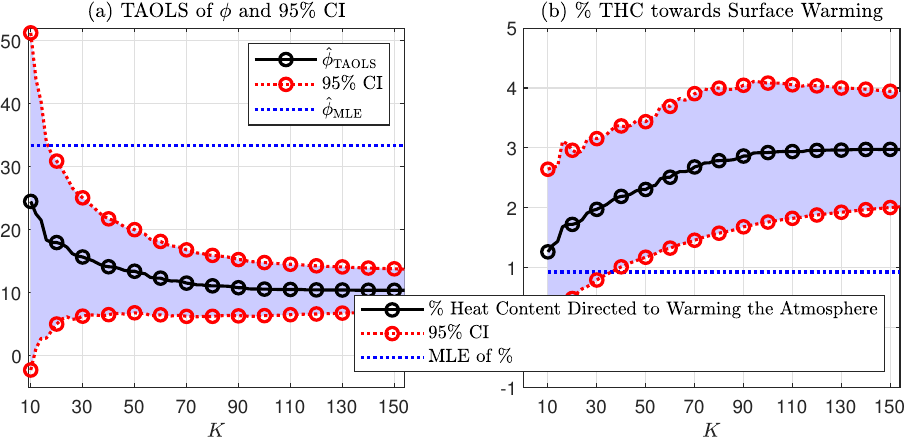}
\caption{TAOLS estimate of $\protect\phi$ and the implied percentage of
total heat content directed toward surface warming in the presence of
multicointegration for different values of $K$ (Model~II). The shaded region
represents the 95\% confidence interval.}
\label{figure:partial_rf_berkeleyt_trendtype_1_mc_phi_percent}
\end{figure}

For Models~I and~III, the estimates of $\phi$ are of comparable magnitude to
those of Model~II. In Model~I (full-efficacy radiative forcing, Berkeley Earth), $\phi$
ranges from $14.69$ to $32.74$ W-yr/m$^{2}$, implying that approximately $%
0.95\%$ to $2.11\%$ of the total heat content contributes to warming the
atmosphere. In Model~III (partial-efficacy radiative forcing, HadCRUT), $\phi$ ranges
from $%
10.82$ to $21.33$ W-yr/m$^{2}$, implying approximately $1.45\%$ to $2.87\%$.
These ranges are broadly consistent with those of Model~II and reinforce the
conclusion that only a small fraction of accumulated heat content drives
surface warming, with the remainder residing predominantly in oceanic
reservoirs.

Following \citet{bruns2020multicointegration}, we plot in Figure~\ref%
{figure:ohc_comparison} the TAOLS-predicted system heat content
against available ocean heat content (OHC) observations for Model~II
(partial-efficacy radiative forcing with Berkeley Earth temperature). The
model-implied cumulative Earth system heat content $Q_{t}$ is converted to
units of $10^{22}$~Joules and scaled to the heat content of the top 2000\,m
of the ocean by multiplying by 0.81, following %
\citet[Section~2.1]{bruns2020multicointegration}. The predicted sequence is
compared with two observational benchmarks: observed ocean heat content in
the top 0--2000\,m and 0--700\,m from \citet{cheng2017improved} and the
simulated series of \citet{marvel2016implications}, scaled to the top
2000\,m by multiplying by 0.88. The TAOLS-predicted OHC series falls between
the 0--700\,m series of \citet{cheng2017improved} and the series of %
\citet{marvel2016implications}, tracking the broad upward trend in ocean
heat uptake since 1940. Results for Models~I and~III are quantitatively
similar to those shown here.

\begin{figure}[ptb]
\centering
\includegraphics[width=0.65%
\textwidth]{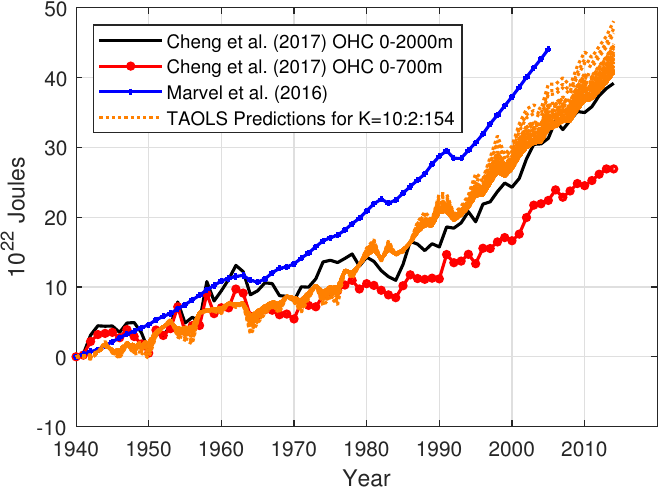}
\caption{TAOLS-predicted system heat content compared with observed ocean
heat content from \citet{cheng2017improved} (top 0--700\,m and 0--2000\,m)
and simulated heat content from \citet{marvel2016implications} for Model~II
(partial-efficacy radiative forcing with Berkeley Earth temperature,
1940--2014). The TAOLS prediction (orange) tracks the 0--2000\,m series of 
\citet{cheng2017improved} and falls between the other two series.}
\label{figure:ohc_comparison}
\end{figure}

\section{Conclusions and Discussion\label{Sec:Conclusions}}

This paper applies the TAOLS estimator of \citet{SPK2025} to the climate
dataset of \citet{bruns2020multicointegration}, providing the first
application of TAOLS to climate data and a direct comparison of the TAOLS
and MLE estimates of ECS. TAOLS is more robust than MLE because it does not
impose Gaussianity on the model innovations and does not require precise
specification of the short-run VAR dynamics---both assumptions that are
difficult to maintain in climate time series that exhibit large non-Gaussian
spikes from volcanic forcing.

Our ECS estimates from TAOLS, which range from $1.81^{\circ}$C to $%
2.49^{\circ}$C across all three data pairings, stand in contrast to higher
estimates, such as those above $5^{\circ}$C from \citet{bjordal2020}, which
emphasize strong positive feedbacks (e.g., cloud dynamics). Our results also
exceed the lowest observational estimates, such as the estimate of
approximately $1.64^{\circ}$C obtained by \citet{lewis2015implications},
which some argue reflects the transient rather than the equilibrium
response. Compared to the MLE from %
\citet{bruns2020multicointegration}, the TAOLS estimates are consistently
lower across all three data pairings. We interpret this gap as evidence that
the Gaussian VAR model underlying the MLE may be misspecified: when the
innovations are non-Gaussian---as they are in climate data with large
volcanic spikes---the MLE-based confidence intervals and point estimates can
be unreliable, whereas TAOLS remains theoretically valid under weaker
distributional assumptions.

The study has several limitations. The multicointegrating structure is
maintained as an assumption, with supporting rank tests drawn directly from %
\citet{bruns2020multicointegration}. All relationships are modeled as
linear, abstracting from potentially nonlinear climate feedbacks~%
\citep{bjordal2020}. The analysis uses global-mean data, setting aside
regional heterogeneity, and the 165-year observational record is short
relative to the multi-century equilibration time of the climate system~%
\citep{bruns2020multicointegration}. As a reduced-form econometric study,
the estimates reflect the statistical properties of the data rather than the
output of a mechanistic simulation.

This study suggests that the equilibrium temperature response to a doubling of CO$_2$ is moderate and remains below $3^{\circ}\mathrm{C}$, although this conclusion should be interpreted in light of the study's limitations. If supported by future work, this result could have implications for climate policy. The analysis also shows that TAOLS can be applied successfully to climate time series. In this application, it produced credible estimates of climate sensitivity with narrower confidence intervals than MLE while avoiding Gaussian distributional assumptions and parametric VAR restrictions. These findings suggest that robust econometric methods may provide a useful alternative for empirical climate research.

\section{Acknowledgments}

The author expresses gratitude to Professor Jingjing Yang at the University of Nevada, Reno and Nie Jiawang at the University of California, San Diego for their guidance and encouragement throughout this project. The author also thanks an anonymous referee for detailed comments and suggestions. 
 \bigskip

\bibliography{Climate_062026}

@misc{berkeleyearth2025data,
  author       = {{Berkeley Earth}},
  title        = {Berkeley Earth Data: Global Temperature and Climate Data},
  howpublished = {\url{https://berkeleyearth.org/data/}},
  note         = {Accessed: 2026-02-07},
  year         = {2025}
}

@article{bjordal2020,
	author = {Bjordal, Jenny and Storelvmo, Trude and Alterskj{\ae}r, Kari and Carlsen, Timo},
	title = {Equilibrium Climate Sensitivity Above 5{$^\circ$}{C} Plausible Due to State-Dependent Cloud Feedback},
	journal = {Nature Geoscience},
	year = {2020},
	volume = {13},
	number = {11},
	pages = {718--721},
	doi = {10.1038/s41561-020-00649-1}
}

@article{cheng2017improved,
	author  = {Cheng, Lijing and Trenberth, Kevin E. and Fasullo, John and Boyer, Tim and Abraham, John and Zhu, Jiang},
	title   = {Improved estimates of ocean heat content from 1960 to 2015},
	journal = {Science Advances},
	year    = {2017},
	volume  = {3},
	number  = {3},
	pages   = {e1601545},
	doi     = {10.1126/sciadv.1601545}
}

@article{marvel2016implications,
	author  = {Marvel, Kate and Schmidt, Gavin A. and Miller, Ron L. and Nazarenko, Larissa S.},
	title   = {Implications for climate sensitivity from the response to individual forcings},
	journal = {Nature Climate Change},
	year    = {2016},
	volume  = {6},
	number  = {4},
	pages   = {386--389},
	doi     = {10.1038/nclimate2888}
}

@article{bruns2020multicointegration,

	title={A multicointegration model of global climate change},

	author={Bruns, Stephan B and Csereklyei, Zsuzsanna and Stern, David I},

	journal={Journal of Econometrics},

	volume={214},

	number={1},

	pages={175--197},

	year={2020},

	publisher={Elsevier},

	doi={10.1016/j.jeconom.2019.05.010}

}

@article{budyko1969effect,
  author  = {Budyko, Mikhail I.},
  title   = {The effect of solar radiation variations on the climate of the {E}arth},
  journal = {Tellus},
  volume  = {21},
  number  = {5},
  pages   = {611--619},
  year    = {1969},
  doi     = {10.3402/tellusa.v21i5.10109}
}

@article{dickey1979distribution,
  author  = {Dickey, David A. and Fuller, Wayne A.},
  title   = {Distribution of the Estimators for Autoregressive Time Series with a Unit Root},
  journal = {Journal of the American Statistical Association},
  volume  = {74},
  number  = {366},
  pages   = {427--431},
  year    = {1979},
  doi     = {10.2307/2286348}
}

@article{EngleGranger1987,
	author = {Engle, R. F. and Granger, C. W. J.},
	title = {Co-Integration and Error Correction: Representation, Estimation, and Testing},
	journal = {Econometrica},
	year = {1987},
	volume = {55},
	number = {2},
	pages = {251--276}
}

@article{GrangerLee1989,
	author = {Granger, C. W. J. and Lee, T.},
	title = {Investigation of production, sales and inventory relationships using multicointegration and non-symmetric error correction models},
	journal = {Journal of Applied Econometrics},
	volume = {4},
	year = {1989},
	pages = {S145--S159},
}

@incollection{GrangerLee1990,
	author = {Granger, C. W. J. and Lee, T.},
	title = {Multicointegration},
	booktitle = {{Advances in Econometrics}},
	editor = {Rhodes, G. F. and Fomby, T. B.},
	volume = {8},
	year = {1990},
	publisher = {JAI Press},
	address = {Greenwich, CT},
	pages = {71--84}
}

@article{gregory2002new,
  author  = {Gregory, Jonathan M. and Stouffer, Ronald J. and Raper, Sarah C. B. and
             Stott, Peter A. and Rayner, Nick A.},
  title   = {An observationally based estimate of the climate sensitivity},
  journal = {Journal of Climate},
  volume  = {15},
  number  = {22},
  pages   = {3117--3121},
  year    = {2002},
  doi     = {10.1175/1520-0442(2002)015<3117:AOBEOT>2.0.CO;2}
}

@article{hillebrand2020editorial,
  author  = {Hillebrand, Enrique and Pretis, Felix and Proietti, Tommaso},
  title   = {Econometric models of climate change: Introduction by the guest editors},
  journal = {Journal of Econometrics},
  volume  = {214},
  number  = {1},
  pages   = {1--5},
  year    = {2020},
  doi     = {10.1016/j.jeconom.2019.09.001}
}

@book{IPCC_2021_WGI,
	address = {Cambridge, UK and New York, NY, USA},
	doi = {10.1017/9781009157896},
	editor = {Masson-Delmotte, V. and Zhai, P. and Pirani, A. and Connors, S. L. and P{\'e}an, C. and Berger, S. and Caud, N. and Chen, Y. and Goldfarb, L. and Gomis, M. I. and Huang, M. and Leitzell, K. and Lonnoy, E. and Matthews, J. B. R. and Maycock, T. K. and Waterfield, T. and Yelek{\c{c}}i, O. and Yu, R. and Zhou, B.},
	pages = {2391},
	publisher = {Cambridge University Press},
	title = {{Climate Change 2021: The Physical Science Basis}},
	url = {https://www.ipcc.ch/report/ar6/wg1/},
	year = {2021}
}

@article{jarque1980efficient,
  author  = {Jarque, Carlos M. and Bera, Anil K.},
  title   = {Efficient Tests for Normality, Homoscedasticity and Serial Independence of Regression Residuals},
  journal = {Economics Letters},
  volume  = {6},
  number  = {3},
  pages   = {255--259},
  year    = {1980},
  doi     = {10.1016/0165-1765(80)90024-5}
}

@article{johansen1992representation,

	title={A representation of vector autoregressive processes integrated of order 2},

	author={Johansen, S{\o}ren},

	journal={Econometric Theory},

	volume={8},

	pages={188--202},

	year={1992},

	publisher={Cambridge University Press}

}

@article{kaufmann1997human,
  author  = {Kaufmann, Robert K. and Stern, David I.},
  title   = {Evidence for human influence on climate from hemispheric temperature relations},
  journal = {Nature},
  volume  = {388},
  pages   = {39--44},
  year    = {1997},
  doi     = {10.1038/40332}
}

@article{kaufmann2010does,
  author  = {Kaufmann, Robert K. and Kauppi, Heikki and Stock, James H.},
  title   = {Does temperature contain a stochastic trend? {E}valuating conflicting statistical results},
  journal = {Climatic Change},
  volume  = {101},
  pages   = {395--405},
  year    = {2010},
  doi     = {10.1007/s10584-009-9711-2}
}

@article{kaufmann2013does,
  author  = {Kaufmann, Robert K. and Kauppi, Heikki and Mann, Michael L. and Stock, James H.},
  title   = {Does temperature contain a stochastic trend? {L}inking statistical results to physical mechanisms},
  journal = {Climatic Change},
  volume  = {118},
  pages   = {729--743},
  year    = {2013},
  doi     = {10.1007/s10584-012-0683-2}
}

@article{kejriwal2013unit,
  author  = {Kejriwal, Mohitosh and Lopez, Carlos},
  title   = {Unit Roots, Level Shifts, and Trend Breaks in Per Capita Output: A Robust Evaluation},
  journal = {Econometric Reviews},
  volume  = {32},
  number  = {8},
  pages   = {892--927},
  year    = {2013},
  doi     = {10.1080/07474938.2012.741063}
}

@article{kwiatkowski1992testing,
  author  = {Kwiatkowski, Denis and Phillips, Peter C. B. and Schmidt, Peter and Shin, Yongcheol},
  title   = {Testing the Null Hypothesis of Stationarity Against the Alternative of a Unit Root},
  journal = {Journal of Econometrics},
  volume  = {54},
  number  = {1--3},
  pages   = {159--178},
  year    = {1992},
  doi     = {10.1016/0304-4076(92)90104-Y}
}

@article{lewis2015implications,
	author    = {Lewis, Nicholas and Curry, Judith A.},
	title     = {The implications for climate sensitivity of {AR5} forcing and heat uptake estimates},
	journal   = {Climate Dynamics},
	year      = {2015},
	volume    = {45},
	number    = {5},
	pages     = {1009--1023},
	doi       = {10.1007/s00382-014-2342-y},
}

@article{myhre1998new,
  author  = {Myhre, Gunnar and Highwood, Eleanor J. and Shine, Keith P. and Stordal, Frode},
  title   = {New estimates of radiative forcing due to well mixed greenhouse gases},
  journal = {Geophysical Research Letters},
  volume  = {25},
  number  = {14},
  pages   = {2715--2718},
  year    = {1998},
  doi     = {10.1029/98GL01908}
}

@misc{noaa2026co2,
  author       = {{National Oceanic and Atmospheric Administration}},
  title        = {Climate change: Atmospheric carbon dioxide},
  howpublished = {\url{https://www.climate.gov/news-features/understanding-climate/climate-change-atmospheric-carbon-dioxide}},
  note         = {Accessed: 2026-02-07},
  year         = {2026}
}

@article{phillips2020transient,
  author  = {Phillips, Peter C. B. and Leirvik, Thomas and Storelvmo, Trude},
  title   = {Econometric estimates of {E}arth's transient climate sensitivity},
  journal = {Journal of Econometrics},
  volume  = {214},
  number  = {1},
  pages   = {6--32},
  year    = {2020},
  doi     = {10.1016/j.jeconom.2019.05.001}
}

@article{PK2024,

	title = {High-dimensional {IV} cointegration estimation and inference},

	journal = {Journal of Econometrics},

	volume = {238},

	number = {2},

	pages = {105622},

	year = {2024},

	issn = {0304-4076},

	doi = {10.1016/j.jeconom.2023.105622},

	url = {https://www.sciencedirect.com/science/article/pii/S030440762300338X},

	author = {Peter C. B. Phillips and Igor L. Kheifets}

}

@article{pretis2020equivalence,
  author  = {Pretis, Felix},
  title   = {Econometric modelling of climate systems: The equivalence of energy balance models and cointegrated vector autoregressions},
  journal = {Journal of Econometrics},
  volume  = {214},
  number  = {1},
  pages   = {256--273},
  year    = {2020},
  doi     = {10.1016/j.jeconom.2019.05.013}
}

@article{RoeBaker2007,
	author = {Gerard H. Roe  and Marcia B. Baker },
	title = {Why Is Climate Sensitivity So Unpredictable?},
	journal = {Science},
	volume = {318},
	number = {5850},
	pages = {629--632},
	year = {2007},
	doi = {10.1126/science.1144735},
	URL = {https://www.science.org/doi/abs/10.1126/science.1144735},
	eprint = {https://www.science.org/doi/pdf/10.1126/science.1144735}}

@article{sellers1969global,
  author  = {Sellers, William D.},
  title   = {A global climatic model based on the energy balance of the earth-atmosphere system},
  journal = {Journal of Applied Meteorology},
  volume  = {8},
  number  = {3},
  pages   = {392--400},
  year    = {1969},
  doi     = {10.1175/1520-0450(1969)008<0392:AGCMBO>2.0.CO;2}
}

@article{SPK2025,
title = {Estimation and inference in a possibly multicointegrated system with a fixed number of instruments},
journal = {Economics Letters},
volume = {250},
pages = {112297},
year = {2025},
issn = {0165-1765},
doi = {10.1016/j.econlet.2025.112297},
author = {Y. Sun and Peter C. B. Phillips and Igor L. Kheifets}
}

@misc{hadcrut4data,
  author       = {{Met Office Hadley Centre and Climatic Research Unit}},
  title        = {{HadCRUT4} Global Surface Temperature Data, Version 4.4.0.0},
  howpublished = {\url{https://www.metoffice.gov.uk/hadobs/hadcrut4/data/4.4.0.0/download.html}},
  year         = {2017},
  note         = {Accessed: 2026}
}

@article{vonschuckmann2023heat,
  author  = {von Schuckmann, Karina and Mini{\`{e}}re, Audrey and Gues, Flora and
             Cuesta-Valero, Francisco Jos{\'e} and Kirchengast, Gottfried and
             Adusumilli, Susheel and Straneo, Fiamma and Ablain, Micha{\"e}l and
             Allan, Richard P. and Barker, Paul M. and others},
  title   = {Heat Stored in the {E}arth System 1960--2020: Where Does the Energy Go?},
  journal = {Earth System Science Data},
  volume  = {15},
  number  = {4},
  pages   = {1675--1709},
  year    = {2023},
  doi     = {10.5194/essd-15-1675-2023}
}

\end{document}